\documentstyle[11pt,aasms4]{article}

\newcommand{\gsim}{\mbox{\hspace{.2em}\raisebox{.5ex}{$>$}\hspace{-.8em}\raisebox{-.5ex}{$\sim$}\hspace{.2em}}}
\newcommand{\lsim}{\mbox{\hspace{.2em}\raisebox{.5ex}{$<$}\hspace{-.8em}\raisebox{-.5ex}{$\sim$}\hspace{.2em}}}
\newcommand{\ssst}{\scriptscriptstyle}
\newcommand{\E}[1]{\times 10^{#1}}
\newcommand{\etal}{et al.}
\newcommand{\lt}{\left}       \newcommand{\rt}{\right}
\newcommand{\RA}[3]{\mbox{R.A.}={#1}^{{\rm h}}{#2}^{{\rm m}}{#3}^{{\rm s}}}
\newcommand{\decl}[3]{\mbox{decl.}={#1}^{\circ}{#2}'{#3}''}

\newcommand{\s}{\,{\rm s}}      \newcommand{\ps}{\,{\rm s}^{-1}}
\newcommand{\yr}{\,{\rm yr}}    \newcommand{\Msun}{M_{\odot}}
\newcommand{\cm}{\,{\rm cm}}    \newcommand{\km}{\,{\rm km}}
\newcommand{\parsec}{\,{\rm pc}}\newcommand{\kpc}{\,{\rm kpc}}
\newcommand{\ergs}{\,{\rm ergs}}        
    \newcommand{\keV}{\,{\rm keV}}

\newcommand{\nel}{n_{e}}        \newcommand{\NH}{N_{\ssst\rm H}}
\newcommand{\no}{n_{\ssst 0}}   \newcommand{\rPDS}{r_{\ssst\rm PDS}}
\newcommand{\vPDS}{v_{\ssst\rm PDS}} \newcommand{\tPDS}{t_{\ssst\rm PDS}}
\newcommand{\Ts}{T_{s}}
\newcommand{\rs}{r_{s}}         \newcommand{\vs}{v_{s}}
\newcommand{\nH}{n_{\ssst\rm H}}        \newcommand{\mH}{m_{\ssst\rm H}}
\newcommand{\nHH}{n({\rm H}_{2})} \newcommand{\NHH}{N({\rm H}_{2})}
\newcommand{\xray}{X-ray}       \newcommand{\Einstein}{{\em Einstein}}
\newcommand{\ROSAT}{{\sl ROSAT}} \newcommand{\ASCA}{{\sl ASCA}}
\newcommand{\du}{d_{8}}         \newcommand{\Eu}{E_{51}}
\newcommand{\lpc}{l_{\rm pc}}   \newcommand{\ru}{r_{3}}
\newcommand{\mE}{\langle E_{\ssst B-V}\rangle}

\begin{document}

\title{\ASCA\ Observations of the Thermal Composite Supernova Remnant 3C~391}

\author{
Yang Chen\altaffilmark{1} and
Patrick O.\ Slane\altaffilmark{2}
}
\altaffiltext{1}{
Department of Astronomy, Nanjing University, Nanjing 210093, P.R.China;
email: ygchen@nju.edu.cn
}
\altaffiltext{2}{
Harvard-Smithsonian Center for Astrophysics, 60 Garden Street,
Cambridge, MA 02138;
email: slane@cfa.harvard.edu
}

\vfil
\begin{abstract}
We present the results from ASCA observations of the centrally enhanced
supernova remnant 3C~391 (G31.9$+$0.0).
We use the \ASCA\ SIS data to carry out an investigation of the spatial
and spectral properties of the X-ray emission from this remnant.
The collisional equilibrium ionization and non-equilibrium ionization
spectral fits indicate that the hot gas within the
remnant has basically reached ionization equilibrium.
The variation of the hydrogen column density across the remnant
is in agreement with the presence of a molecular cloud to the northwest.
The comparisons of
hydrogen column and \xray\ hardness between the NW and SE portions
of the remnant support a scenario in which the SNR has broken out of
a dense region into an adjacent region of lower density.
The mean density within the SNR is observed to be much lower
than the immediate ambient cloud density. This and the centrally brightened
\xray\ morphology can be explained either by the evaporation of
engulfed cloudlets or by a radiative stage of evolution for the remnant.

\keywords{
 radiation mechanisms: thermal ---
 supernova remnants: individual: 3C~391 (G31.9+0.0) ---
 X-rays: ISM
}

\end{abstract}

\section{Introduction}

Apart from the well-known classification of
shell-like, Crab-like, and composite supernova remnants (SNRs),
a collection of SNRs with shell-like radio emission and centrally brightened
thermal X-rays has attracted substantial attention
(e.g., White \& Long 1991 [=WL91];
Rho \& Petre 1998; Wilner, Reynolds, \& Moffett 1998).
Many of these so-called ``thermal composite'' or ``mixed morphology''
remnants (such as W28, W44, IC~443, G349.7+0.2, and 3C~391) are found to produce
the hydroxyl radical 1720 MHz maser emission that is characteristic of a shock
interaction with dense molecular gas (Green \etal\ 1997).

3C~391 (G31.9+0.0) is similar to other remnants in this
collection. Its X-ray brightness peaks well inside the radio
shell, and the spectrum is that of a thermal plasma. The remnant
shows an elongation from northwest (NW) to southeast (SE) in the
radio band, and the centroid of the soft X-ray emission lies in
the SE region. Reynolds \& Moffett (1993) (= RM93) first pointed
out that the radio and \xray\ morphologies can be explained with a
gas breakout from a molecular cloud into a lower density region.
In a \ROSAT\ \xray\ study, Rho \& Petre (1996) (= RP96) could not
actually distinguish between an increase in hydrogen column
density or a decrease in temperature across the remnant from SE to
NW from the spectra alone, and favored the hydrogen column
variation based on the morphology. Frail et al. (1996) suggest
that OH masers in the direction of 3C~391 are associated with the
remnant, and Reach \& Rho (1996) find strong enhancement in [OI]
emission near the northwestern edge of the remnant, which is
indicative of shock interaction. CO and other molecular line
observations confirm the location of the remnant at the
southwestern edge of a molecular cloud (Wilner \etal\ 1998; Reach
\& Rho 1999). The far-infrared H$_{2}$O and OH emission lines are
consistent with the passage of shock wave through dense clumps
(Reach \& Rho 1998), and  broad CS and CO line sources are found
to coincide with one of the OH maser regions (Reach \& Rho 1999).

Because of the limited spectral resolution of the \ROSAT\ PSPC
(0.5-2.2 keV) and \Einstein\ IPC ($<4\keV$), the previous \xray\
instruments did not resolve line features in the SNR spectrum.
\ASCA, with its broader energy range (0.5-10 keV) and better
energy response, allows us to analyze the line emission, element
abundances, and the narrow band images to investigate the physics
of the hot gas inside 3C~391 and inspect the breakout model more
fully.

\section{Data Analysis}
The \ASCA\ observation of 3C~391 was made on 21-22 April 1994 with both
the solid-state imaging spectrometer (SIS) and the gas imaging
spectrometer (GIS).
In the SIS observation, both 1-CCD and 4-CCD clocking modes were used.
The observational data were screened with the standard ``rev2'' criteria
(see \ASCA\ ABC guide).
The SIS0 and SIS1 bright mode data were exploited in this analysis due to
the better spectral and spatial resolution of the SIS data over the GIS data.
The relevant observation parameters are listed in Table 1a.

\subsection{Spectral Analysis}\label{sec:spec}
The spectra for the whole remnant (shown in Figure~1) were extracted from
a circular region of a radius 3.5~arcmin,
centered at $\RA{18}{49}{28}.3$, $\decl{-00}{56}{15}$ (J2000)
(see the diagram in Figure~3{\em a} below).
Background spectra were extracted from a nearby source-free field.
The net count rates of the two CCD mode SIS0 and SIS1 spectra
are tabulated in Table~1b.
Since data of different clocking modes have different spectral responses,
we would not merge the spectra but fit them simultaneously to the same model.

In order to search for spectral variations across the remnant,
we also extracted spectra from two circular regions of
radius 1.7~arcmin in the NW and SE halves (as diagramed in Figure~3{\em a}).
The net count rates of these spectra are also tabulated in Table~1b.
All the spectra mentioned above were regrouped to contain at least
25 net counts per bin.

There are three prominent line features in the spectra.
To determine accurately the energy of the lines,
following Bamba \etal\ (2000), we first fitted the spectra of the whole
remnant to a thermal bremsstrahlung and three Gaussian lines with a
Morrison \& McCammon (1983) interstellar absorption.
The best-fit line centers are $1.35\pm0.01$, $1.85\pm0.01$,
and $2.46\pm0.02\keV$ (the errors are 90\% confidence).
They correspond to Mg He$\alpha$, Si~He$\alpha$, and S~He$\alpha$
emission lines.
The Fe L complex at 1-$1.5\keV$ could be present in the spectra.
The emission diminishes rapidly above the photon energy $\sim 4\keV$
and no Fe~K$\alpha$ emission is observed,
indicating that the gas temperature is not very high assuming
normal abundances (below $1\keV$).
We used an absorbed collisional equilibrium ionization (CEI) model,
VMEKAL (Mewe, Kaastra, \& Liedahl 1995), in the XSPEC code to
fit the two clocking mode SIS0 and SIS1 spectra of the whole remnant
simultaneously.
In the spectral fitting, the abundances of the elements (Mg, Si, \& S)
showing evident emission lines, are treated as free parameters,
while other element abundances are fixed at the default solar values.
The spectral fit for the whole remnant are shown in Figure~1.
Similar fit procedures are also applied to the spectra of
the NW and SE portions individually.
We find no significant contribution of a high energy tail that might be
associated with a strong nonthermal component to the spectrum.
The CEI fit results are summarized in Table 2.

On the other hand, we also fit these spectra to VNEI, a non-equilibrium
ionization (NEI) model, in the XSPEC11.0 code.
The NEI fit results are presented in Table 3.
The goodness-of-fit and most of the fit parameters are
 similar for the CEI and NEI cases, and
the ionization parameter $\nel t$ obtained from the NEI
models is $>10^{12}\cm^{-3}\s$,
so the hot gas in the remnant is basically in
ionization equilibrium.

The metal abundances yielded with both the CEI and NEI models (see Tables
2 \& 3) are essentially consistent with solar values and thus seem to be
consistent with an interstellar composition.
From the two tables we can not see significant difference in temperature
between the SE and NW regions of 3C~391 in either model,
but find different error ranges of $\NH$ for the two regions
(with higher $\NH$ in the NW than in the SE).
To check this, we compute the two-dimensional error contours for
$\NH$ and $kT$ for the SE and NW regions (Figure~2).
In both the CEI and NEI cases, the confidence contours for the two regions
are clearly disjoint, so $kT$ and $\NH$ highly correlated for each region.
The contours show similar range of temperature but different range of
column density for the two regions.
We also fit the NW and SE spectra simultaneously to the same temperature
and the same column density, respectively, using a CEI model;
here the metal abundances for the two regions given in Table~2 are used.
The best-fit results are summerized in Table~4.
For the case of the same temperature, one would again see that the
one dimensional error ranges of $\NH$ for the two regions do not overlap.
Assuming the same column density, on the other hand, leads to another
possible case in which the NW region is higher in temperature but lower
in emission measure than the SE.
In the latter case, considering the immediate adjacency of the molecular cloud
to the northwest of the remnant (Wilner \etal\ 1998),
it is quite unlikely that the gas density the NW region is lower than
that in the SE region,
so the hot gas in the defined circular NW region should actually occupy
larger volume than in the SE region.
Though this possibility can not be eliminated, a variation in column density
between the SE and NW region, ie., an increase in column density from
the SE to the NW is consistent with the proximity of the molecular cloud.

\subsection{Spatial Analysis}\label{sec:img}
The SIS0 observation covers the entire remnant,
but the SIS1 observation misses a small part of the eastern rim.
We have thus used only SIS0 data for the image production.
The corrections for exposure and vignetting were made
for the two clocking modes (1-CCD and 4-CCD) separately, and
an adjustment was made to the reconstructed pointing direction
to correct for documented erroneous star-tracker readings (Gotthelf
et al.\ 2000).
The hard (2.6-10$\keV$) emission gray-scale image is overlaid with
the soft (0.5-2.6$\keV$) emission contours in Figure~3{\em a}.
The ``hard'' emission was extracted above the $2.6\keV$
so as to not include the lines of
Mg He$\alpha$, Si He$\alpha$, and S He$\alpha$
and can basically be regarded as thermal continuum.
In Figure~3{\em b} and Figure~3{\em c}, the 1.5 GHz radio contours
(from Moffett \& Reynolds 1994) are overlaid with the
soft and hard emission images, respectively.
We also present the
narrow band Mg He$\alpha$ (1.2-1.5$\keV$), Si He$\alpha$
(1.7-2.0$\keV$), and S He$\alpha$ (2.3-2.6$\keV$) images
in Figure~3{\em d}, Figure~3{\em e}, and Figure~3{\em f}.
The three energy ranges (ER1, ER2, and ER3), in which the narrow band
emissions are extracted, are labeled in Figure~1.
The three narrow band images contain 2308, 2587, and 833 counts,
individually.
Note that the three images are not corrected for the
underlying continuum contribution
and the Mg band (ER1) also includes emission from the Fe~L blend.
The Mg, Si, and S emissions have 56\%, 55\%, and 36\% energy
contributions, respectively, in each energy range.
All these image maps have been adaptively smoothed so as to contain a
minimum of 50 counts in the top hat filter.
The two OH maser points (Frail et al.\ 1996) are labeled in
the maps.

\section{Discussion}
\subsection{Absorbing hydrogen}
Both the CEI and NEI models give a hydrogen column density $\NH$ around
$3.0\E{22}\cm^{-2}$ (Tables 2 \& 3).
In the spectral fits above, we show disjoint error ranges of $\NH$
for the NW and SE regions of 3C~391,
and favor a variation in $\NH$ between the two regions.
This is consistent with the northwestward enhancement of CO emission
(Wilner \etal\ 1998) and with the southeastward breakout
scenario (RM93).
In the CEI model fit, the best-fit value of $\NH$ for the NW portion is
$\sim4\E{21}\cm^{-2}$ higher than that for the SE portion;
and in the NEI model fit, this difference is $\sim6\E{21}\cm^{-2}$.
If the difference of $\NH$ for the two portions reflects the
density contrast between the inside and outside of the
molecular cloud,
the molecular column inside the cloud would be
$\NHH\sim3$-$4\E{20}\cm^{-2}$
(following RM93, considering that the photoionization cross section
for ${\rm H}_{2}$ at $1\keV$, per H atom,
is about about 8 times that for atomic hydrogen [Brown \& Gould 1970]).
In this way the mean molecular density inside the cloud would be
$\langle\nHH\rangle\sim100\lpc^{-1}\cm^{-3}$,
where $\lpc$ is the depth in pc of the molecular layer.
This would imply $\langle\nHH\rangle$ of order $\sim10$-$20\cm^{-3}$
if the line-of-sight depth of the SNR in the cloud is
similar to the remnant radius (6-9 pc).

\subsection{Images}
The soft band map (Fig.3a,b) is very similar to the \ROSAT\ PSPC image (RP96).
The hard \xray\ image (Fig.3a,c) looks relatively bright in the NW
compared with that in the SE,
in contrast to the soft \xray\ image which looks faint in the NW.
This fact can be explained by increased extinction from the NW cloud,
and is consistent with a model in which the gas in the SE region
has broken out into a lower density environment.
There seems to be a narrow \xray\ bridge-like structure
at the center of the \xray\ images connecting the SE and NW portion,
which might possibly represents the ``nozzle" or ``tunnel" of
the NW-to-SE breakout.

The emission dominated by Mg and Si (Fig.3d,e) is centrally brightened
in the SE half but faint in the NW (similar to the soft band map),
implying obscuration in the NW half.
An arc-like feature emerges in the NW of the S He$\alpha$ emission map
(Fig.3f), while it is absent in the Mg emission map (Fig.3d)
and marginally evident in the Si emission map (Fig.3e).
This is consistent with an absorption effect in the softer
Mg and Si lines.
This arc-like shaped feature is close to the NW radio shell
and reveals a hot gas structure behind the blastwave and
embedded in the dense cloud.

In each \xray\ map the SE half is centrally brightened.
The hard emission (continuum) in the NW is also centrally brightened,
and is conspicuously faintest along the western limb where the radio
brightness is a maximum.
The mechanism
responsible for the enhanced central emission
may be interior cloudlet evaporation (WL91)
or, perhaps, the cooling of the rim gas (e.g.\ Harrus et al. 1997,
Rho \& Petre 1998, Cox et al. 1999, Shelton et al. 1999).
We discuss these two mechanisms in \S\ref{sec:dyn}).

The peak of the X-ray brightness distribution is located at
about $\RA{18}{49}{33}.5$, $\decl{-00}{56}{37}$ (J2000)
for both the soft map and the narrow band maps,
and the hard map peaks at about
$\RA{18}{49}{25}$, $\decl{-00}{54}{32}$ (J2000).
The association of 3C~391 with a dense molecular cloud
is suggestive of a massive progenitor star.
The gravitational core collapse of the massive progenitor
should have left behind a compact star,
such as that observed for IC~443 (Olbert et al.\ 2001),
another ``thermal composite''
which is similar to 3C~391 in many aspects.
While the broad spatial response of ASCA prohibits a sensitive
search for an embedded compact star, future {\sl Chandra} and
{\sl XMM} observations near
the peaks in the soft and hard maps mentioned above are clearly
of interest.

\subsection{Distance}
The presence of the HI absorption against 3C~391 out to the tangent point
velocity $\sim+105\km\ps$ puts the remnant beyond $7.2\kpc$ (for a
Galactocentric radius $8.5\kpc$),
and the absence of the absorption at negative velocities sets an
upper limit of $11.4\kpc$
(Caswell \etal\ 1971; Radhakrishnan \etal\ 1972; RM93).
This range is supported by the discovery of two OH 1720 MHz maser
features (at +105 and $+110\km\ps$) in the direction of 3C~391 (Wilner,
Reynolds, \& Moffett 1998).
With the hydrogen column density ($\NH\sim3\E{22}\cm^{-2}$)
obtained from the \xray\ spectral fits,
one can estimate another upper limit to the distance.
The extinction per unit distance within $2\kpc$ in the direction of 3C~391
is $\mE/d\sim0.60\,{\rm mag}\kpc^{-1}$ (Lucke 1978).
The extinction beyond $2\kpc$, in the direction of the Galactic center,
should be higher than this value.
The correlation $\NH=5.9\E{21}\mE\cm^{-2}$
(Spitzer 1978; Predehl \& Schmitt 1995) then
gives $d\lsim8.5\kpc$,
which is in agreement with the above range of distance.
In this paper $d\sim8\du\kpc$ will be used hereafter.

\subsection{\xray\ emitting gas and its dynamics}\label{sec:dyn}
If the unabsorbed flux in the VMEKAL model (case CEI) is adopted,
the \xray\ (0.5-10 keV) luminosity is $L_{x}\sim2.7\E{36}\du^{2}\ergs\ps$.
The best-fit volume emission measure (EM) of the remnant
in the CEI case is $f\nel\nH V \sim 1.0\E{59}\du^{2}\cm^{-3}$,
where $f$ is the filling factor of the \xray\ emitting plasma;
the EM value in the NEI case is a bit smaller.
Based on the VLA observation (RM93),
we approximate the remnant volume as a cylinder of a diameter
$5'$ and a height $7'$.
With $\nel\approx1.2\nH$ assumed, the emission measure yields
$\nH\sim1.5f^{-1/2}\du^{-1/2}\cm^{-3}$.
The \xray\ emitting mass is
$M_{x}=1.4\nH\mH fV\sim92f^{1/2}\du^{5/2}\Msun$,
which indicates that the emission is
dominated by swept-up ambient matter.
(The NEI case would correspond to 0.86 times the above
gas density and mass.)
The spectral fits show a trend
of larger EM of the NW region than that of the SE region.
This may imply that the NW part of the remnant gas
is denser than the SE part.

The mean interior hot gas density $\nH$ is much smaller than the environment
H-atom density $2\langle\nHH\rangle$ inside the molecular cloud,
indicating that a large amount of ambient matter does not act as
\xray\ emitting gas after being swept up or
engulfed by the supernova blastwave.
The explanation could be either that we are observing the hot, tenuous
internal gas while the dense material near the rim has cooled down,
or that the medium inside the cloud is clumpy and
most of the \xray\ emitting interclump medium (ICM) was
evaporated from the clumps.
The two mechanisms can both lead to the observed centrally brightened
\xray\ morphology, as we discuss below.

\subsubsection{Cloud evaporation case}
Here we use the self-similar solution incorporating cloud evaporation
(WL91) to discuss the dynamics of SNR 3C~391.
Because cloud evaporation slowly increases the interior density, the mean
interior hot gas density could be a few times the postshock density.
This gas can provide the centrally emitting thermal X-ray component
observed ``mixed morphology'' remnants such as 3C~391 and others
(Harrus et al. 1997, Harrus et al. 2001).
The postshock temperature $\Ts$ can be derived from
the observed \xray\ emitting gas temperature $kT_{x}$($\sim0.52\keV$
in the CEI case)
using a scaling factor
\begin{equation}
q=KT_{x}/1.27\Ts,
\end{equation}
where $q$ and the energy ratio constant $K$ (scaled by Sedov value)
are dependent on $C/\tau$ (WL91).
Here $C$ is the ratio of the mass in the clumps to the mass of ICM,
and $\tau$ is the ratio of the cloud evaporation time to the SNR's age.
We follow RP96 and take $C/\tau\sim3$-5.
The velocity of the blastwave can be obtained from
$\vs=(16k\Ts/3\mu\mH)^{1/2}$
where the mean atomic weight $\mu=0.61$,
which then gives the dynamical age of the remnant
$t=2r_{s}/5\vs$.
Since the SNR is complicated in morphology, we scale the
radius with a mean value ($3'$) of the whole radio volume: $\rs=3'\ru$.
The hot gas density distribution is dependent upon the model parameters;
for simplicity, however, we assume the mean interior density is
twice the postshock density, which is broadly consistent with a range
of profiles for the cloud evaporation model.
Thus the undisturbed preshock ICM density $\no$ is about
$\nH/8\sim0.2f^{-1/2}\du^{-1/2}\cm^{-3}$.
This density can also be estimated from the \xray\ luminosity
$L_{x}$ ($\sim2.7\E{36}\du^{2}\ergs\ps$) using WL91's
eq.(21) and it is thus in a consistent range
$\sim0.1$-$0.2\ru^{-3/2}\du^{-1/2}$.
The explosion energy is given by
\begin{equation}
E=\frac{16\pi(1.4\no\mH)}{25(\gamma+1)K}\frac{r_{s}^{5}}{t^{2}},
\end{equation}
where the adiabatic index $\gamma=5/3$. The results of the
dynamical parameters obtained from the above relations are
tabulated in Table~5. The explosion energy (1.3-3.4)$\E{50}\ergs$
is somewhat lower than the canonical value of $10^{51}\ergs$. The
age estimate in RP96's evaporation model ($\sim6$-$8\E{3}\yr$) is
higher than that obtained here ($\sim4$-$5\E{3}\ru\du\yr$),
majorly because they used a larger-than-average radius of remnant.
RM93 used a Sedov model with a high preshock density, so their
estimate of age is rather large ($\sim1.7\E{4}\yr$).

\subsubsection{Radiative rim case}
In this case we assume the clumpy mass is not important and
the mean molecular density $\langle\nHH\rangle$ would correspond to
a uniform undisturbed hydrogen density $\no\sim30\cm^{-3}$.
The radius of the SNR at the beginning of the radiative
pressure-driven snowplow (PDS) stage is given by
(Cioffi, McKee, \& Bertschinger 1988)
\begin{equation}
\rPDS=14.0\Eu^{2/7}\no^{-3/7}\zeta_{m}^{-1/7}\parsec,
\end{equation}
where $\zeta_{m}$ is the metallicity factor and is close to unity
for normal abundances. Adopting an explosion energy $\Eu\equiv
E/(10^{51}\ergs)\sim1$, we have $\rPDS\sim3.3\parsec$, which is
smaller than the remnant's radius $\rs\gsim6\du\parsec$ (here a
radius $2.5'$ of the NW radio shell is adopted). Thus, we expect
at least the NW part of the SNR, which appears to be in contact
with the dense cloud, to have already approached the PDS phase. In
fact, the \xray\ emission in the NW does not extend out to the
radio shell as it does in the SE. This is similar to what is
observed in W44, in which the remnant has entered the radiative
phase, shutting down the X-ray at the radio shell (Harrus et al.\
1997). The newly detected near-infrared [Fe II] and the
mid-infrared 12-18 $\mu$m emission reveals the radiative shell of
3C~391, particularly in the NW rim (Reach, Rho, \& Jarrett 2001).
We thus assume that the shell of 3C~391 has cooled sufficiently to
have reached the radiative phase. In this case the hot interior of
the remnant drives the cooled shell to expand and is responsible
for the centrally enhanced cooling (e.g.\ Cioffi et al.\ 1988).
Although ignored here, the effects of thermal conduction enhance
this process through smoothing of the interior temperature profile
(Cox et al. 1999, Shelton et al. 1999).

The postshock temperature should be lower than that for
the Sedov case: $k\Ts<0.77kT_{x}\sim0.4\keV$;
the shock velocity $\vs$ is then slower than $\sim580\km\ps$.
In the PDS stage, $\rs$ and $\vs$ follow the formulae
(Cioffi et al.\ 1988)
\begin{eqnarray}
\rs &=& \rPDS\lt(\frac{4t}{3\tPDS}-\frac{1}{3}\rt)^{3/10}, \\
\vs &=& \vPDS\lt(\frac{4t}{3\tPDS}-\frac{1}{3}\rt)^{-7/10},
\end{eqnarray}
where
$\vPDS = 413\no^{1/7}\zeta_{m}^{3/14}\Eu^{1/14}\km\ps$ and
$\tPDS = 1.33\E{4}\Eu^{3/14}\no^{-4/7}\zeta_{m}^{-5/14}\yr$.
From the above, one has
\begin{equation}
\Eu = \lt(\frac{\rs}{14\parsec}\rt)^{98/31}
    \lt(\frac{\vs}{413\km\ps}\rt)^{42/31}
    \no^{36/31}\zeta_{m}^{5/31}, \\
\end{equation}
which yields $\Eu<9.2(\ru\du)^{98/31}$,
which is reasonable although it offers a weak overall constraint.
From Cioffi et al.\ (1988), the age of the remnant can readily be
obtained:
\begin{equation}
t=3.3\E{3}\lt(\frac{\rs}{14\parsec}\rt)
          \lt(\frac{\vs}{413\km\ps}\rt)^{-1}
          \lt[3 + \no^{-10/31} \zeta_{m}^{-10/31}
          \lt(\frac{\rs}{14\parsec}\rt)^{-10/31}
          \lt(\frac{\vs}{413\km\ps}\rt)^{40/31} \rt]\yr
\end{equation}
which yields $t>4.3\E{3}\ru\du\yr$.
If we use $\Eu$ as a parameter, then the present shock velocity is
$\vs\sim110\Eu^{31/42}(\ru\du)^{-7/3}\km\ps$
and the remnant age is
$t\sim1.9\E{4}\Eu^{-31/42}(\ru\du)^{10/3}\yr$.

The cloud evaporation and the radiative rim mechanisms
both produce acceptable dynamical parameters.
A final judgment between them may depend on higher
resolution observations with {\sl Chandra} and {\sl XMM-Newton} which
may, for example, yield fine radial brightness and temperature profile or
reveal smaller scale features that might support
the clumpy ISM scenario.

\section{Conclusion}
We have investigated the spatial and spectral properties of the SNR 3C~391
using \ASCA\ SIS data.
The CEI and NEI spectral fits indicate that the hot gas within the
SNR has basically reached ionization equilibrium.
The hydrogen column density in the direction of the NW portion of
the SNR is higher than that of the SE portion,
in agreement with the location of the molecular cloud to the NW.
The comparison of the hydrogen column
and \xray\ hardness between the NW and SE portions
supports the NW-to-SE breakout scenario, as suggested by earlier
observations of this SNR.
The much lower mean density within the SNR than the immediate
ambient cloud density and the centrally brightened \xray\ morphology
can be explained either by an SNR evolving in a clumpy cloud
inside of which gas is evaporated from the engulfed cloudlets,
or by an SNR which has entered the radiative stage with the interior
gas still hot, but with the rim material cooled down.

\acknowledgements{
The authors would like to
thank Steve Reynolds and David Moffett for providing the radio
image of 3C~391.
We also thank Randall Smith for helpful discussions
related to this paper.
A special gratitude should be ascribed to an anonymous referee
whose comments help to improve the manuscript appreciably.
Part of YC's work was carried out in the CfA.
YC acknowledges support from CNSF grant 1007003
and grant NKBRSF-G19990754 of China Ministry of Science and Technology.
POS acknowledges support from NASA contract NAS8-39073 and grant
NAG5-9281.
}

\clearpage

\centerline{\begin{tabular}{rccr}
\multicolumn{4}{c}{Table 1a: Summary of observations (used)} \\ \hline\hline
Seq.\ No. & Instrument & clocking mode & exposure \\ \hline
ad51017000s000102h & SIS0 & 4-CCD &     36 s \\
          s000202h & SIS0 & 4-CCD & 10,479 s \\
          s000302m & SIS0 & 1-CCD & 23,927 s \\
          s000502m & SIS0 & 4-CCD &     26 s \\
          s000602m & SIS0 & 1-CCD &     94 s \\
ad51017000s100102h & SIS1 & 4-CCD & 10,594 s \\
          s100202m & SIS1 & 1-CCD & 24,056 s \\
          s100402m & SIS1 & 4-CCD &     26 s \\ \hline
\end{tabular}}
\bigskip

\centerline{\begin{tabular}{cccc}
\multicolumn{4}{c}{Table 1b: Summary of net count rates (in ${\rm counts}\ps$)
} \\ \hline\hline
CCD chips & whole & NW & SE \\ \hline
SIS0 1-CCD & $(18.7\pm0.3)\E{-2}$ & $(5.3\pm0.2)\E{-2}$ & $(7.2\pm0.2)\E{-2}$ \\
SIS1 1-CCD & $(15.2\pm0.3)\E{-2}$ & $(4.9\pm0.1)\E{-2}$ & $(5.8\pm0.2)\E{-2}$ \\
SIS0 4-CCD & $(17.3\pm0.4)\E{-2}$ & $(5.4\pm0.2)\E{-2}$ & $(6.3\pm0.3)\E{-2}$ \\
SIS1 4-CCD  & $(13.0\pm0.4)\E{-2}$ & $(4.3\pm0.2)\E{-2}$ & $(4.5\pm0.2)\E{-2}$ \\ \hline
\end{tabular}}
\clearpage

\centerline{\begin{tabular}{c|ccc}
\multicolumn{4}{c}{Table 2: VMEKAL (CEI) fitting results with the 90\%
confidence ranges}\\ \hline\hline
                & whole & NW & SE \\ \hline
$f\nel\nH V/\du^{2}$ ($10^{58}\cm^{-3}$) & $10.39^{+1.95}_{-1.75}$ &
        $4.16^{+1.46}_{-1.20}$ & $3.71^{+1.11}_{-0.93}$\\
$kT_{x}$ (keV) & $0.52\pm0.03$ & $0.52^{+0.05}_{-0.04}$ & $0.51\pm0.03$\\
{[Mg/H]} & $1.36^{+0.18}_{-0.17}$ & $1.16^{+0.35}_{-0.30}$ &
        $1.50^{+0.31}_{-0.27}$ \\
{[Si/H]} & $1.00\pm0.10$ & $0.81^{+0.17}_{-0.14}$ &
        $1.07^{+0.17}_{-0.19}$ \\
{[S/H]} & $0.67\pm0.14$ & $0.53^{+0.23}_{-0.21}$ &
        $0.63^{+0.26}_{-0.22}$ \\
$\NH$ ($10^{22}\cm^{-2}$) & $3.03^{+0.10}_{-0.11}$ &
        $3.34^{+0.17}_{-0.18}$ & $2.96\pm0.14$ \\
$F(0.5$-$10\keV)$ ($\ergs\cm^{-2}\ps$) & $4.7\E{-12}$ & $1.5\E{-12}$ &
        $1.7\E{-12}$\\
$F^{(0)}(0.5$-$10\keV)$ ($\ergs\cm^{-2}\ps$) & $3.5\E{-10}$ &
        $1.4\E{-10}$ & $1.2\E{-10}$\\
$\chi^{2}/{\rm d.o.f.}$ & $349/230$ & $125/108$ & $201/118$ \\ \hline
\multicolumn{4}{l}{Here $f$ denotes the filling factor of the hot gas,
$F$ the absorbed flux,}\\
\multicolumn{4}{l}{and $F^{(0)}$ the unabsorbed flux.} \\
\end{tabular}}
\bigskip

\centerline{\begin{tabular}{c|ccc}
\multicolumn{4}{c}{Table 3: VNEI fitting results with the 90\%
confidence ranges}\\ \hline\hline
                & whole & NW & SE \\ \hline
$f\nel\nH V/\du^{2}$ ($10^{58}\cm^{-3}$) & $7.38^{+1.46}_{-1.09}$ &
        $3.50^{+1.08}_{-0.88}$ & $1.99^{+0.90}_{-0.44}$\\
$kT_{x}$ (keV) & $0.59^{+0.03}_{-0.01}$ & $0.57^{+0.04}_{-0.05}$ &
        $0.63^{+0.07}_{-0.03}$\\
$n_{e}t$ ($10^{13}\cm^{-3}\s$) & $0.85^{+\infty}_{-0.10}$ &
        $3.86^{+\infty}_{-0.52}$ & $3.76^{+\infty}_{-0.41}$ \\
{[Mg/H]} & $1.37^{+0.23}_{-0.19}$ & $1.25^{+0.46}_{-0.33}$ &
        $1.64^{+0.21}_{-0.40}$ \\
{[Si/H]} & $1.22^{+0.15}_{-0.12}$ & $1.02^{+0.23}_{-0.19}$ &
        $1.38^{+0.27}_{-0.24}$ \\
{[S/H]} & $0.75^{+0.17}_{-0.16}$ & $0.59^{+0.29}_{-0.26}$ &
        $0.68^{+0.32}_{-0.29}$ \\
$\NH$ ($10^{22}\cm^{-2}$) & $2.90\pm0.07$ &
        $3.33\pm0.15$ & $2.70^{+0.16}_{-0.06}$ \\
$F(0.5$-$10\keV)$ ($\ergs\cm^{-2}\ps$) & $4.8\E{-12}$ & $1.5\E{-12}$ &
        $1.8\E{-12}$ \\
$F^{(0)}(0.5$-$10\keV)$ ($\ergs\cm^{-2}\ps$) & $2.3\E{-10}$ &
        $1.1\E{-10}$ & $6.4\E{-11}$ \\
$\chi^{2}/{\rm d.o.f.}$ & $359/229$ & $133/107$ & $201/117$ \\ \hline
\multicolumn{4}{l}{Here $f$ denotes the filling factor of the hot gas,
$F$ the absorbed flux,}\\
\multicolumn{4}{l}{and $F^{(0)}$ the unabsorbed flux.} \\
\end{tabular}}
\bigskip
\clearpage

\centerline{\begin{tabular}{c|ccc}
\multicolumn{3}{c}{Table~4: Results of simultaneous VMEKAL fitting}\\
\multicolumn{3}{c}
{for NW and SE, assuming the same $kT_{x}$}\\
\multicolumn{3}{c}{and the same $\NH$, respectively.}\\ \hline\hline
                & NW & SE \\ \hline
$kT_{x}$ (keV) & \multicolumn{2}{c}{$0.51^{+0.04}_{-0.02}$}\\
$\NH$ ($10^{22}\cm^{-2}$) & $3.36\pm0.12$ & $2.95^{+0.10}_{-0.11}$\\
$f\nel\nH V/\du^{2}$ ($10^{58}\cm^{-3}$)
        & $4.35^{+0.84}_{-0.75}$ & $3.60^{+0.67}_{-0.61}$\\
$\chi^{2}/{\rm d.o.f.}$ & \multicolumn{2}{c}{$326/233$} \\ \hline
$\NH$ ($10^{22}\cm^{-2}$) & \multicolumn{2}{c}{$3.13^{+0.08}_{-0.09}$}\\
$kT_{x}$ (keV) & $0.57\pm0.03$ & $0.48\pm0.02$\\
$f\nel\nH V/\du^{2}$ ($10^{58}\cm^{-3}$)
        & $3.00^{+0.49}_{-0.37}$ & $4.82^{+0.73}_{-0.68}$\\
$\chi^{2}/{\rm d.o.f.}$ & \multicolumn{2}{c}{$337/233$} \\ \hline
\end{tabular}}
\vspace{15mm}

\centerline{\begin{tabular}{c|c}
\multicolumn{2}{c}{Table 5: Dynamical parameters for the}\\
\multicolumn{2}{c}{evaporation model of WL91}\\ \hline\hline
$C/\tau^{\dag}$ & 3-5\\
$K^{\ddag}$ & 0.385-0.189\\
$q^{\ddag}$ & 0.372-0.143\\
$k\Ts$ (keV) & 0.42-0.54\\
$\vs$ ($\km\ps$) & 590-670 \\
$r_{s}$ (pc)  & $7.0\ru\du$ \\
$t$ ($10^{3}\yr$) & (4.6-4.0)$\ru\du$ \\
$E$ ($10^{50}\ergs$) & (1.3-3.4)$f^{-1/2}\ru^{3}\du^{5/2}$ \\ \hline
\multicolumn{2}{l}{$\dag$ following RP96} \\
\multicolumn{2}{l}{$\ddag$ adopted from WL91} \\
\end{tabular}}
\clearpage

\begin{center}
\section{Figure captions}
\end{center}

\figcaption[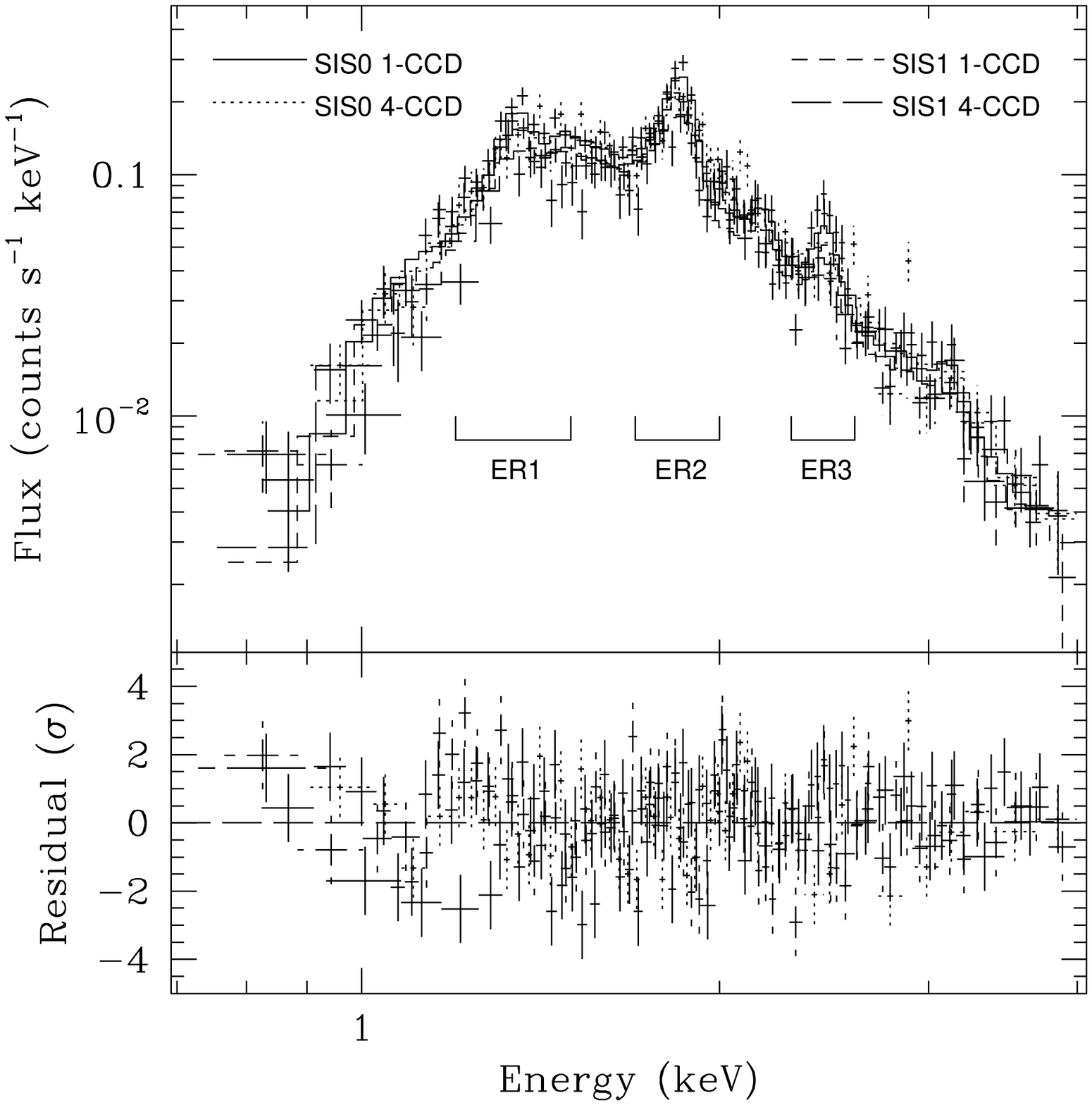]{
The SIS \xray\ spectra of 3C391 fitted with the VMEKAL model.
The three energy ranges (ER) are labelled, in which the narrow band
images are produced.
}

\figcaption[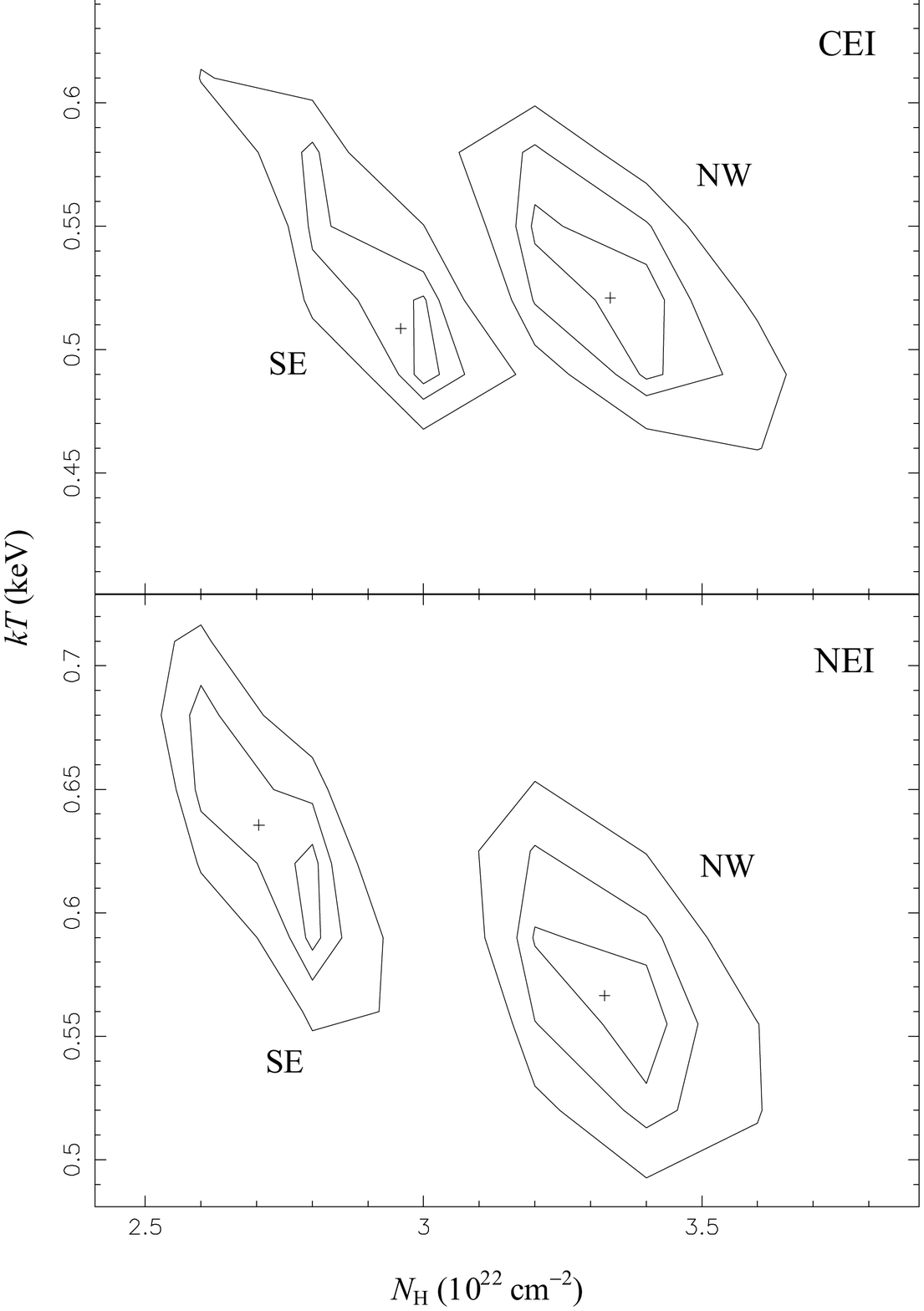]{
Two dimensional confidence contours for the NW and SE regions of 3C~391.
The contours from inner to outer correspond to $\Delta\chi^{2}=$2.30 (68\%),
4.61 (90\%), and 9.21 (99\%).
}

\figcaption[f3.ps]{
The \xray\ images produced using SIS0 data.
Panel {\em a} is the hard (2.6-10$\keV$) emission gray-scale image
overlaid with the soft (0.5-2.6$\keV$) emission contours.
The three circles in panel {\em a} designate the areas
from which the SIS spectra are extracted.
Panels {\em b} and {\em c} display the gray-scale images of
soft and hard emission, respectively, overlaid with the dashed contours
of 1.5 GHz radio emission (from Moffett \& Reynolds [1994]).
Panels {\em d}, {\em e}, and {\em f} are the narrow band
gray-scale images and solid contours of
Mg He$\alpha$ (1.2-1.5$\keV$),
Si He$\alpha$ (1.7-2.0$\keV$),
and S He$\alpha$ (2.3-2.6$\keV$) emissions.
In panel {\em f} the image is also overlaid with the dashed radio contours.
The seven levels of solid contours are linear between the maximum
and 20\% maximum brightness.
The two cross labels in each panel denote the OH maser points
(Frail et al.\ 1996).
}

\clearpage
\begin{thebibliography}{}
\bibitem{} Bamba, A., Yokogawa, J., Sakano, M., \& Koyama, K. 2000,
PASJ, 52, 259
\bibitem{} Brown, R.L., \& Gould, R.J. 1970, Phys.\ Rev.\ D, 1, 2252
\bibitem{} Caswell, J.L., Dulk, G.A., Goss, W.M., Radhakrishnan, V.,\&
Green, A.J. 1971, A\&A, 12, 271
\bibitem{} Cioffi, D.F., McKee, C.F., \& Bertschinger, E. 1988, ApJ,
334, 252
\bibitem{} Cox, D. P., Shelton, R. L., Maciejewski, W., Smith, R. K.,
Plewa, T., Pawl, A., \& R\'{o}zyczka, M. 1999, ApJ, 524, 179
\bibitem{} Frail, D.A., Goss, W.M., Reynoso, E.M., Giacani, E.B.,
Green, A.J., \& Otrupcek, R. 1996, AJ, 111, 1651
\bibitem{} Gotthelf, E. V., Ueda, Y., Fujimoto, R., Kii, T., \& Yamaoka, K.
2000, ApJ, 543, 417
\bibitem{} Green, A. J., Frail, D. A., Goss, W. M., Otrupcek, R.
1997, AJ, 114, 2058
\bibitem{} Harrus, I. M., Hughes, J. P, Singh, K. P., Koyama, K.,
\& Asaoka, I. 1997, ApJ, 488, 781
\bibitem{} Harrus, I. M., Slane, P. O., Smith, R. K., \& Hughes, J. P.
2001, ApJ, in press
\bibitem{} Mewe, R., Kaastra, J.S., \& Liedahl, D.A. 1995, Legacy, 6, 16
\bibitem{} Moffett, D. A., \& Reynolds, S. P., 1994, ApJ, 425, 668
\bibitem{} Morrison, R., \& McCammon, D., 1983, ApJ, 270, 119
\bibitem{} Olbert, C.M.,  Clearfield, C.R., Williams, N.E., Keohane, J.W.,
\& Frail, D.A. 2001, ApJ Letter, in press (astro-ph/0103268)
\bibitem{} Predehl, P., \& Schmitt, J.H.M.M. 1995, A\&A, 293, 889
\bibitem{} Radhakrishnan, V., Goss, W.M., Murray, J.D., \& Brooks, J.W.
1972, ApJS, 24, 49
\bibitem{} Reach, W.T., \& Rho, J.H., 1996 A\&A, 315, L277
\bibitem{} ------------------, 1998, ApJ, 507, L93
\bibitem{} ------------------, 1999, ApJ, 511, 836
\bibitem{} Reach, W.T., Rho, J.H., \& Jarrett, T.H., 2001, astro-ph/0108173
\bibitem{} Reynolds, S.P., \& Moffett, D.A. 1993, AJ, 105, 2226 (RM93)
\bibitem{} Rho, J.H., \& Petre, R. 1996, ApJ, 467, 698 (RP96)
\bibitem{} Rho, J.H., \& Petre, R. 1998, ApJ, 503, L167
\bibitem{} Shelton, R.  L., Cox, D. P., Maciejewski, W., Smith, R. K.,
Plewa, T.; Pawl, A., \& R\'{o}zyczka, M. 1999, ApJ, 524, 192
\bibitem{} Spitzer, L. Jr. 1978, Physical Processes in the Interstellar
Medium (Wiley, New York)
\bibitem{} Wilner, D.J., Reynolds, S.P., \& Moffett, D.A. 1998, ApJ, 115, 247
\bibitem{} White, R.L., \& Long, K.S. 1991, ApJ, 373, 543 (WL91)
\end{thebibliography}
\end{document}